\documentclass[a4paper,twocolumn,11pt,accepted=2024-02-06]{quantumarticle}
\pdfoutput=1
\usepackage[utf8]{inputenc}
\usepackage[english]{babel}
\usepackage[T1]{fontenc}
\usepackage{amsmath}
\usepackage{hyperref}
\usepackage{orcidlink}
\usepackage{tikz}
\usepackage{lipsum}

\begin{document}

\title{Quantitative relations between different measurement

contexts} 
\author{Ming Ji \orcidlink{0000-0002-6569-5099}}
\email{physmji@gmail.com}
\affiliation{Graduate School of Advanced Science and Engineering, Hiroshima University, Kagamiyama 1-3-1, Higashi Hiroshima 739-8530, Japan}
\author{Holger F. Hofmann \orcidlink{0000-0001-5649-9718}}
\email{hofmann@hiroshima-u.ac.jp}
\affiliation{Graduate School of Advanced Science and Engineering, Hiroshima University, Kagamiyama 1-3-1, Higashi Hiroshima 739-8530, Japan}
\maketitle

\begin{abstract}
    
  In quantum theory, a measurement context is defined by an orthogonal basis in a Hilbert space, where each basis vector represents a specific measurement outcome. The precise quantitative relation between two different measurement contexts can thus be characterized by the inner products of nonorthogonal states in that Hilbert space. Here, we use measurement outcomes that are shared by different contexts to derive specific quantitative relations between the inner products of the Hilbert space vectors that represent the different contexts. It is shown that the probabilities that describe the paradoxes of quantum contextuality can be derived from a very small number of inner products, revealing details of the fundamental relations between measurement contexts that go beyond a basic violation of noncontextual limits. The application of our analysis to a product space of two systems reveals that the nonlocality of quantum entanglement can be traced back to a local inner product representing the relation between measurement contexts in only one system. Our results thus indicate that the essential nonclassical features of quantum mechanics can be traced back to the fundamental difference between quantum superpositions and classical alternatives.
  
\end{abstract}

\section{Introduction}
\label{sec:I}
Quantum contextuality is a characteristic feature of quantum mechanics that makes it impossible to assign predetermined context-independent values to the outcomes of a set of different measurements that cannot be performed jointly. As shown by the Bell-Kochen-Specker theorem~\cite{Bel64,Koc67}, the formal proof of contextuality can be given by comparing the predictions of noncontextual models with the statistical predictions of quantum theory. These contradictions can then be characterized by violations of statistical limits imposed by such noncontextual models~\cite{Cab08,Bad09,Kle12,Pan13,Su15,Kun15,Xu16,Kri17,Kun18,Sch18,Lei20}.
Perhaps the best-known example are Bell inequalities, which also demonstrate that the relations between contexts can be nonlocal~\cite{Bel66,Har92,Har93,Bos97,Gen05,Zel07,Car18,Tem19}. However, the failure of noncontextual models is not sufficient to explain the physics of quantum contextuality. All by itself, the violation of an inequality only tells us what quantum theory is not. The real issue that still needs to be resolved is the precise nature of contextual relations between the statistics of different measurements. This paper examines how the Hilbert space formalism replaces noncontextual logic with quantitative relations between the statistics of incompatible measurements. By translating the classical noncontextual relations used to derive inequalities directly into relations between Hilbert space inner products, we show in detail how quantum mechanics modifies the relations between different measurement contexts. It is then possible to gain deeper insights into the microscopic structure of the relations between the outcomes obtained in incompatible measurements.

A specific measurement context is described by a set of distinct measurement outcomes represented by orthogonal Hilbert space vectors. Measurement outcomes described by nonorthogonal Hilbert space vectors do not share the same measurement context. In a noncontextual framework, the relation between these outcomes is undefined. This is in stark contrast with quantum theory, where the relation between measurement outcomes that do not share any measurement context is uniquely defined by the inner product of their Hilbert space vectors. Here, we consider the deterministic conditions that the relations between different contexts impose on these inner products. In Hilbert spaces of three or more dimensions, the same measurement outcome can belong to two or more contexts, establishing relations between probabilities of measurement outcomes in different contexts that can be used to derive Hardy-like proofs of quantum contextuality~\cite{Cab13}. In such Hardy-like proofs of quantum contextuality, the initial state is usually defined by its orthogonality relations with other mutually nonorthogonal states, a feature that uniquely defines the states that maximally violate noncontextual assumptions~\cite{Ji23}. In quantum theory, the relations between different contexts defined by shared measurement outcomes uniquely determine the relations between inner products of Hilbert space vectors representing the outcomes that can only be obtained in different measurement contexts. In this sense, the Hilbert space relations expressed by inner products represent fundamental quantitative relations between different measurement contexts that define the difference between classical noncontextual logic and quantum mechanics. Here, we investigate how contextuality relations based on shared measurement outcomes of different contexts relate different inner products of Hilbert space vectors to each other, thus characterizing the quantitative relation of different contexts, that can be utilized as a tool to directly evaluate contextuality. Our starting point is the relation between measurement contexts where one outcome is shared and another outcome of one context is given by a quantum superposition of two outcomes in the other context. We can then derive a relation between two outcomes from different contexts mediated by a central context connected to the contexts of these two outcomes by one shared outcome each. This fundamental relation between different contexts can be expressed by a factorization of the inner product between the initial two states into a pair of inner products relating the two states with the central context. Combined with the normalization of Hilbert space vectors, this method allows us to derive more complicated relations between Hilbert space inner products in arbitrarily large networks of interrelated contexts. It is therefore possible to derive the well-known contextuality paradoxes from a small set of elementary relations expressing the characteristic features of quantum contextuality in their most compact form.

In the following, we apply our method to construct the basic cyclic structure of interrelated contexts needed for Hardy-like violations of noncontextuality. In this cyclic structure, the initial two outcomes $|D1\rangle$ and $|D2\rangle$ belonging to two different contexts are related to each other by a central context and by an alternative sequence of two contexts centered on a specific outcome $|f\rangle$. In the cyclic structure formed by these two different relations between $|D1\rangle$ and $|D2\rangle$, noncontextual logic would seem to require that the outcome $|f\rangle$ can only be obtained when either $|D1\rangle$ or $|D2\rangle$ are possible as well. This is the basis of a Hardy-like contextuality paradox, since the Hilbert space vector $|N_{\mathrm f}\rangle$ that is orthogonal to both $|D1\rangle$ and $|D2\rangle$ has a nonvanishing inner product with the outcome $|f\rangle$. By applying our method, we can derive the inner product between $|N_{\mathrm f}\rangle$ and $|f\rangle$ from the initial inner products between $|D1\rangle$, $|D2\rangle$ and an additional outcome $|3\rangle$  based on the relations between contexts that have a shared measurement outcome. In this derivation, the precise relation between $|f\rangle$ and $|N_{\mathrm f}\rangle$ is determined from a combination of two different contextual relations between $|D1\rangle$ and $|D2\rangle$, one of which is mediated by the outcome $|3\rangle$ and another that is mediated by the outcome $|f\rangle$. The emergence of a paradoxical relation between different measurement contexts can then be traced to a more fundamental set of relations between inner products that do not appear to be paradoxical when viewed in isolation. In principle, contexts with three outcomes each are sufficient for the construction of a sufficiently complex network. However, the result is not limited to three dimensional Hilbert spaces. By applying our result to a product space of two systems, we can derive the original Hardy paradox and demonstrate that the well-established relation between nonlocality and contextuality is based on the same fundamental network of measurement contexts. It is then possible to identify the essential features of quantum nonlocality by focusing on the relation between contexts defined by only three possible measurement outcomes. This result may help to illustrate the merit of identifying the most elementary quantitative relations between different measurement contexts. It is very likely that all manifestations of quantum contextuality can be broken down into smaller pieces, where the essential relations can be represented by chains of Hilbert space relations corresponding directly to the cyclic relations between contexts used to formulate the specific scenario. Our method of analysis thus simplifies the analysis of quantum nonlocality by greatly reducing the number of outcomes that needs to be considered in higher dimensional Hilbert spaces.

The rest of the paper is structured as follows. In Sec.~\ref{sec:II} we show how the relations between different measurement contexts defined by shared outcomes can be used to derive quantitative relations between Hilbert space inner products. In Sec.~\ref{sec:III} we construct cyclic relations between outcomes from different measurement contexts and obtain a quantitative measure for the most basic Hardy-like violation of noncontextuality. In Sec.~\ref{sec:IV}, we show that our analysis can be applied to a product Hilbert space of two systems and derive the necessary violation of locality predicted by quantum theory. Sec.~\ref{sec:V} summarizes the results and concludes the paper.

\section{Quantitative relations between different Hilbert space inner products}
\label{sec:II}

In quantum mechanics, each measurement context is described by an orthogonal set of Hilbert space vectors that represent the possible measurement outcomes. If the vectors describing two measurement outcomes are not orthogonal, these measurement outcomes cannot share the same measurement context. It should therefore be possible to describe the relation between different measurement contexts by the inner products between the Hilbert space vectors that represent measurement outcomes from different contexts. The more familiar meaning of these inner products is given by Born's rule, where one of the two state vectors represents the initial state and the absolute value square of the inner product gives the probability of obtaining the measurement result represented by the other state vector. In order to relate different measurement outcomes with each other, we can thus consider a situation where a specific measurement outcome can be obtained with certainty. In this case, the initial state is represented by the same Hilbert space vector that also represents the measurement outcome. Thus, the quantum formalism is based on a fundamental symmetry between the description of initial states and the description of measurement outcomes, and relations between different measurements can be described using the relation between state preparation and measurement expressed by quantum superpositions and Born's rule. We can always start from a prediction with a certainty of 100$\%$ for a measurement outcome in one context and then use the inner product between this outcome and any measurement outcome in another context to predict the probability of that outcome. Quantum contextuality is different from noncontextual expectations since there is no other possible probability for such an outcome, even though the only condition for the initial state was the certainty of an outcome in a seemingly unrelated context. It is therefore possible to derive relations between different measurement contexts based on the relations between inner products of Hilbert space vectors representing the outcomes of different measurement contexts.

\begin{figure}[t]
\begin{picture}(500,200)
\put(50,0){\makebox(125,200){
\scalebox{0.55}[0.55]{
\includegraphics{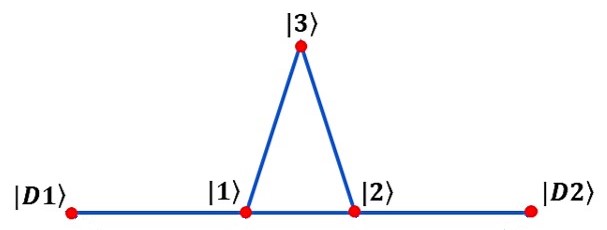}}}}
\end{picture}
\caption{\label{fig1}
Diagram of the measurement context $\{|1\rangle$, $|2\rangle$, $|3\rangle\}$ and the outcomes $|D1\rangle$ and $|D2\rangle$ belonging to other measurement contexts. Each node represents a measurement outcome and each line between the nodes represents an orthogonality relation between the different measurement outcomes.
}
\end{figure}

In general, measurement outcomes can belong to the same measurement context if the states that represent them are orthogonal to each other. These orthogonality relations can be represented by graphs where a line between different outcomes indicates that the two outcomes are orthogonal to each other~\cite{Bud22}. The shared measurement outcomes between different contexts define the directly observable relations between different measurement contexts. Although the number of different outcomes corresponds to the dimensionality of a given Hilbert space, only two outcomes need to be exchanged to define a different context. The strongest contextuality relations can therefore be obtained by considering only three different outcomes between the contexts.
We thus start our analysis from the measurement outcomes represented by three mutually orthogonal states $|1\rangle$, $|2\rangle$, and $|3\rangle$. Next, we consider an outcome $|D1\rangle$ that is orthogonal to $|1\rangle$, but not to $|2\rangle$ or $|3\rangle$. This means that $|D1\rangle$ shares a context with $|1\rangle$, but not with $|2\rangle$ or $|3\rangle$. We also assume that any additional outcome that shares a context with $|1\rangle$, $|2\rangle$, and $|3\rangle$ also shares a context with $|D1\rangle$. This means that the outcome $|D1\rangle$ can be expressed by a quantum superposition of the two measurement outcomes $|2\rangle$ and $|3\rangle$,
\begin{equation}
\label{eq:D1}
|D1\rangle=|2\rangle\langle2|D1\rangle+|3\rangle\langle3|D1\rangle.
\end{equation}
In this manner, quantum mechanics defines a nontrivial contextuality relation between measurement outcomes from different contexts. If there is a sequence of contextuality relations, it is possible to use the superposition expressed by Eq.~(\ref{eq:D1}) to derive quantitative relations between different inner products of measurement outcomes. A basic relation between different inner products can be obtained by introducing an outcome $|D2\rangle$ that is orthogonal to $|2\rangle$, but not to $|1\rangle$ or $|3\rangle$. The corresponding superposition is then given by
\begin{equation}
\label{eq:D2}
|D2\rangle=|1\rangle\langle1|D2\rangle+|3\rangle\langle3|D2\rangle.
\end{equation}
The outcomes $|D1\rangle$ and $|D2\rangle$ are connected by the central context $\{|1\rangle$, $|2\rangle$, $|3\rangle\}$ as shown in Fig.~\ref{fig1}. $|D1\rangle$ and $|D2\rangle$ do not share any context since they have a nonzero inner product. When this inner product is determined using the superpositions given by Eq.~(\ref{eq:D1}) and Eq.~(\ref{eq:D2}), the result reads
\begin{equation}
\label{eq:D1D2}
\langle D1|D2\rangle=\langle D1|3\rangle\langle3|D2\rangle.
\end{equation}
The separate relations of $|D1\rangle$ and $|D2\rangle$ with the context $\{|1\rangle$, $|2\rangle$, $|3\rangle\}$ can thus be combined to determine the relation between the outcomes $|D1\rangle$ and $|D2\rangle$. Using Born's rule to interpret this relation, the probability of finding $|D2\rangle$ when the initial state predicts the outcome $|D1\rangle$ with certainty is equal to the product of the probability of finding the outcome $|3\rangle$ for an initial state of $|D1\rangle$ times the probability of finding $|D2\rangle$ for an initial state of $|3\rangle$. This corresponds to a sequence of a measurement projecting $|D1\rangle$ into $|3\rangle$, followed by a projection of $|3\rangle$ into $|D2\rangle$. It would seem as if the projection of the state onto the intermediate result $|3\rangle$ had no effect on the final outcome $|D1\rangle$ because a measurement of $|D1\rangle$ in $|D2\rangle$ required that the state was $|3\rangle$ all along. This is in fact the fundamental assumption behind the construction of contextuality paradoxes using pre- and post-selection~\cite{Lei05}.

It should be noted that the result of Eq.~(\ref{eq:D1D2}) expresses an important aspect of the relation between noncontextual logic and quantum contextuality. The ``classical'' interpretation of Eq.~(\ref{eq:D1D2}) based on a joint reality of $|D1\rangle$ and $|3\rangle$ conditioned by $|D2\rangle$ does not work because Hilbert space inner products cannot be interpreted as classical alternatives. Instead, Eq.~(\ref{eq:D1D2}) descibes a nontrivial quantitative relation between three measurement outcomes that do not share any measurement contexts and are therefore never observed jointly. The following analysis is based on the observation that quantitative relations such as Eq.~(\ref{eq:D1D2}) can be obtained from the relations between measurement contexts defined by their shared measurement outcomes as shown in Fig.~\ref{fig1}. Although this relation does not appear to contradict noncontextual logic, we can extend this method of analysis to construct alternative connections between the same two outcomes $|D1\rangle$ and $|D2\rangle$ resulting in a cyclic relation that needs to be self-consistent. In the following, we do this by introducing additional measurement contexts.

\begin{figure}[t]
\begin{picture}(500,200)
\put(50,0){\makebox(135,200){
\scalebox{0.55}[0.55]{
\includegraphics{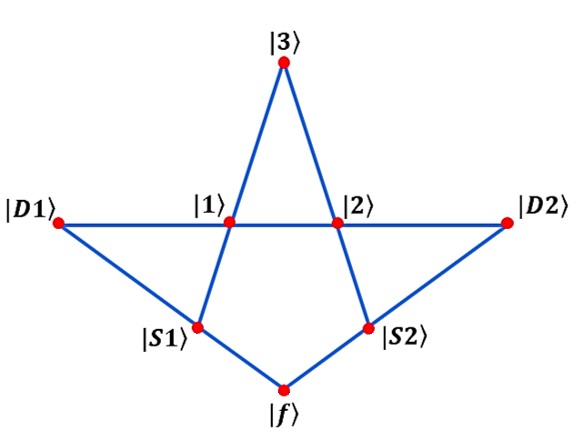}}}}
\end{picture}
\caption{\label{fig2}
Diagram of the measurement contexts $\{|1\rangle, |2\rangle, |3\rangle\}$, $\{|1\rangle, |D1\rangle, |S1\rangle\}$, $\{|2\rangle, |D2\rangle, |S2\rangle\}$ and the additional outcome $|f \rangle$ sharing a context with both $|S1\rangle$ and $|S2\rangle$. Each node represents a measurement outcome and each line between two nodes represents an orthogonality relation between the different measurement outcomes.
}
\end{figure}

In order to find other contextuality relations, we need to introduce additional states that share a context with $|D1\rangle$ and $|D2\rangle$, respectively. We can do that by introducing the states $|S1\rangle$ and $|S2\rangle$ that are part of the contexts $\{|1\rangle, |D1\rangle, |S1\rangle\}$ and $\{|2\rangle, |D2\rangle, |S2\rangle\}$. Similar to $|D1\rangle$ ($|D2\rangle$), the additional outcome $|S1\rangle$ ($|S2\rangle$) can also be expressed as a superposition of the outcomes $|2\rangle$ and $|3\rangle$ ($|1\rangle$ and $|3\rangle$) because it is assumed that any additional outcome added to the context $\{|1\rangle, |2\rangle, |3\rangle\}$ can also be added to the context $\{|1\rangle, |D1\rangle, |S1\rangle\}$ ($\{|2\rangle, |D2\rangle, |S2\rangle\}$). As a result of this relation with the central context, $|S1\rangle$ and $|S2\rangle$ are not orthogonal and cannot share the same context. In order to find a second relation between $|D1\rangle$ and $|D2\rangle$, it is therefore necessary to introduce a third additional measurement outcome $|f\rangle$ that shares a context with both $|S1\rangle$ and $|S2\rangle$. As before, we also assume that $|f\rangle$ shares a context with any state that can be added to the context $\{|1\rangle, |2\rangle, |3\rangle\}$, ensuring that $|f\rangle$ can be expressed by a superposition of $|1 \rangle$, $|2\rangle$, and $|3\rangle$,
\begin{eqnarray}
\label{eq:f1}
|f\rangle=|1\rangle\langle1|f\rangle+|2\rangle\langle2|f\rangle+|3\rangle\langle3|f\rangle.
\end{eqnarray}
The additional outcomes and their orthogonality relations with each other are shown in Fig.~\ref{fig2}. It is now possible to relate the outcomes $|D1\rangle$ and $|D2\rangle$ to each other via the outcome $|f\rangle$. Due to its orthogonality with $|S1\rangle$ and $|S2 \rangle$, $|f\rangle$ can be expressed by superpositions that include either $|D1\rangle$ or $|D2\rangle$,
\begin{eqnarray}
\label{eq:f2f3}
|f\rangle&=&|1\rangle\langle1|f\rangle+|D1\rangle\langle D1|f\rangle, \nonumber\\
|f\rangle&=&|2\rangle\langle2|f\rangle+|D2\rangle\langle D2|f\rangle.
\end{eqnarray}
Note that these superpositions seem to suggest that the outcome $|f\rangle$ is only obtained when the system is either in $|D1\rangle$ or in $|D2 \rangle$, since the ``options'' $|1\rangle$ and $|2\rangle$ are orthogonal to each other and seem to exclude each other for that reason. However, such reasoning is not consistent with the mathematical form of superpositions in quantum mechanics, since the direct application of a logical ``either/or" is a misinterpretation of the relation expressed by a quantum superposition. In the present case, we can obtain a more useful relation by subtracting Eq.~(\ref{eq:f1}) from the sum of the two equations in Eq.~(\ref{eq:f2f3}). The result reads
\begin{equation}
\label{eq:f4}
|f\rangle=|D1\rangle\langle D1|f\rangle+|D2\rangle\langle D2|f\rangle-|3\rangle\langle3|f\rangle.
\end{equation}
The significance of this relation is not as clear as the basic relation between inner products shown in Eq.~(\ref{eq:D1D2}). The outcome $|f\rangle$ is expressed as a quantum superposition of outcomes from different contexts, establishing a relation between $|D1\rangle$ and $|D2\rangle$ that depends on the relation between the outcomes $|f\rangle$ and $|3\rangle$.

It may be interesting to compare the superposition in Eq.~(\ref{eq:f4}) with the noncontextual logic commonly applied to construct Hardy-like contextuality paradoxes~\cite{Cab13}. The key difference between quantum superpositions and noncontextual models is the assumption that contextuality relations should be interpreted in terms of logical alternatives. If the outcome $|f\rangle$ is obtained, $|D1\rangle$ or $|1\rangle$ and $|D2\rangle$ or $|2\rangle$ are the only available alternatives. Since $|1\rangle$ and $|2\rangle$ contradict each other, all possibilities must include either $|D1\rangle$ or $|D2\rangle$. A naive application of this noncontextual expectation to quantum superpositions would suggest that the state $|f\rangle$ could be expressed by a superposition of the two nonorthogonal states $|D1\rangle$ and $|D2\rangle$. Although Eq.~ (\ref{eq:f4}) is not sufficient to prove that this is not the case, the appearance of an additional contribution associated with the outcome $|3\rangle$ shows that quantum contextuality modifies the logic that relates different contexts to each other. In the next section, we will show that the outcome $|f\rangle$ must have a component orthogonal to both $|D1\rangle$ and $|D2\rangle$ by introducing an additional measurement outcome that represents this component. It is then possible to derive the fundamental structure of Hardy-like contextuality paradoxes using the relations between $|D1\rangle$ and $|D2\rangle$ given by Eqs.~(\ref{eq:D1D2}) and (\ref{eq:f4}).

\section{Analysis of quantum contextuality}
\label{sec:III}

In the previous section, we derived relations between the measurement outcomes $|D1\rangle$ and $|D2\rangle$ through their inner products with measurement outcomes in other contexts. It is then possible to identify quantitative relations between the different measurement outcomes. These quantitative relations are different from the relations defined by noncontextual logic because quantum superpositions do not express logical alternatives. Based on noncontextual logical alternatives, the outcome $|f\rangle$ should only be obtained if the initial state included either $|D1\rangle$ or $|D2\rangle$. This expectation can be represented by an additional measurement outcome $|N_{\mathrm f}\rangle$ that shares a context with both $|D1\rangle$ and $|D2\rangle$. $|N_{\mathrm f}\rangle$ should also share a context with any outcome that can only occur if the initial state includes either $|D1\rangle$ or $|D2\rangle$ since there is no conceivable situation where such an outcome could be obtained jointly with $|N_{\mathrm f}\rangle$. In order to establish the link between our quantitative analysis of inner products and the characteristic paradoxes of quantum contextuality, we thus need to consider an additional outcome $|N_{\mathrm f} \rangle$ that shares a context with both $|D1\rangle$ and $|D2\rangle$. It is then possible to use the relations of quantum contextuality given by Eq.~(\ref{eq:D1D2}) and Eq.~(\ref{eq:f4}) to derive the inner product between the outcome $|N_{\mathrm f}\rangle$ and the outcome $|f\rangle$, allowing us to trace the violation of noncontextual logic back to the nonvanishing inner products that characterize the relation between the outcomes $|D1\rangle$ and $|D2\rangle$.

The introduction of the measurement outcome $|N_{\mathrm f}\rangle$ is different from the measurement outcomes shown in Fig. \ref{fig2} because it shares a measurement context with both $|D1\rangle$ and $|D2\rangle$. Mathematically, it is therefore defined by the orthogonality relations
\begin{eqnarray}
\label{eq:Nfdef}
\langle D1|N_{\mathrm f}\rangle &=& 0,
\nonumber \\
\langle D2|N_{\mathrm f}\rangle &=& 0.
\end{eqnarray}
Consistent with the previous definitions, we shall also require that any outcome added to the context $\{|1\rangle, |2\rangle, |3\rangle\}$ shares a context with $|N_{\mathrm f}\rangle$, so that $|N_{\mathrm f}\rangle$ can be expressed as a superposition of these three states,
\begin{eqnarray}
\label{eq:Nf}
|N_{\mathrm f}\rangle=|1\rangle\langle1|N_{\mathrm f}\rangle+|2\rangle\langle2|N_{\mathrm f}\rangle+|3\rangle\langle3|N_{\mathrm f}\rangle.
\end{eqnarray}
The definition of the outcome $|N_{\mathrm f}\rangle$ given by Eq.~(\ref{eq:Nfdef}) can be used to determine the inner product with the outcome $|f\rangle$ based on the relation in Eq.~(\ref{eq:f4}). The result shows that $|N_{\mathrm f}\rangle$ and $|f\rangle$ are related via the outcome $|3\rangle$,
\begin{equation}
\label{eq:Nfnegative}
\langle f|N_{\mathrm f}\rangle=-\langle f|3\rangle\langle3|N_{\mathrm f}\rangle,
\end{equation}
It is worth noting that this relation looks very similar to the relation between $|D1\rangle$ and $|D2\rangle$ given by Eq.~(\ref{eq:D1D2}). However, the right hand side of the equation is negative, and neither $|f\rangle$ nor $|N_{\mathrm f}\rangle$ share a context with either $|1\rangle$ or $|2\rangle$. It is therefore less obvious what kind of relation between $|f\rangle$ and $|N_{\mathrm f}\rangle$ Eq.~(\ref{eq:Nfnegative}) describes.

To obtain a precise relation between $|f\rangle$ and $|N_{\mathrm f}\rangle$, it is necessary to relate the expansion of the state $|N_{\mathrm f}\rangle$ in Eq.~(\ref{eq:Nf}) to the inner products $\langle D1|3\rangle$ and $\langle D2|3\rangle$. This can be done using the orthogonality relations of Eq.~(\ref{eq:Nfdef}) and the fact that $|D1\rangle$ and $|D2\rangle$ share a context with the states $|1\rangle$ and $|2\rangle$, respectively. The relations between the inner products are given by
\begin{eqnarray}
\label{eq:Nf2}
\langle D2|1\rangle\langle1|N_{\mathrm f}\rangle&=&-\langle D2|3\rangle\langle3|N_{\mathrm f}\rangle, \nonumber\\
\langle D1|2\rangle\langle2|N_{\mathrm f}\rangle&=&-\langle D1|3\rangle\langle3|N_{\mathrm f}\rangle.
\end{eqnarray}
We can thus obtain a set of relations between the expansion coefficients in Eq.~(\ref{eq:Nf}). Using the normalization of the states $|D1\rangle$ and $|D2\rangle$, it is possible to express these relation in terms of the inner products $\langle D1|3\rangle$ and $\langle D2|3\rangle$,
\begin{eqnarray}
\label{eq:squareNf}
|\langle1|N_{\mathrm f}\rangle|^{2}&=&\frac{|\langle D2|3\rangle|^{2}}{1-|\langle D2|3\rangle|^{2}}|\langle3|N_{\mathrm f}\rangle|^{2},
\nonumber \\
|\langle2|N_{\mathrm f}\rangle|^{2}&=&\frac{|\langle D1|3\rangle|^{2}}{1-|\langle D1|3\rangle|^{2}}|\langle3|N_{\mathrm f}\rangle|^{2}.
\end{eqnarray}
Due to the normalization of the state $|N_{\mathrm f}\rangle$, these relations can be used to determine the absolute values of all three inner products in the expansion given by Eq.~(\ref{eq:Nf}). Specifically, we can now show that the inner products that determine the relation between $|D1\rangle$ and $|D2\rangle$ in Eq.~(\ref{eq:D1D2}) also determine the inner product between $|N_{\mathrm f}\rangle$ and $|3\rangle$ that appears in Eq.~(\ref{eq:Nfnegative}),
\begin{equation}
\label{eq:3Nf}
|\langle 3|N_{\mathrm f}\rangle|^{2}=\frac{(1-|\langle D1|3\rangle|^{2})(1-|\langle D2|3\rangle|^{2})}{1-|\langle D1|3\rangle|^{2}|\langle D2|3\rangle|^{2}}.
\end{equation}
This result has important implications for the contextuality relations between the outcomes. Since the inner product of $|N_{\mathrm f}\rangle$ and $|3\rangle$ is necessarily nonzero, $|N_{\mathrm f}\rangle$ and $|3\rangle$ cannot share the same context. With respect to Eq.~(\ref{eq:f4}), it is important to note that the outcome $|3\rangle$ cannot be expressed as a superposition of only $|D1\rangle$ and $|D2\rangle$, since it would then have to be orthogonal to $|N_{\mathrm f}\rangle$. Eq.~(\ref{eq:f4}) thus shows that $|f\rangle$ cannot be expressed as a superposition of only $|D1\rangle$ and $|D2\rangle$ either, resulting in the corresponding violation of noncontextual logic.

According to Eq.~(\ref{eq:Nfnegative}), we still need to determine the inner product $\langle f|3\rangle$ in order to find the inner product of $|N_{\mathrm f} \rangle$ and $|f\rangle$ that describes the contradiction between quantum contextuality and noncontextual logic. We note that the relation between $|f\rangle$ and $|3\rangle$ in the contextuality diagram of Fig.~\ref{fig2} is similar to the relation between $|D1\rangle$ and $|D2\rangle$, where the central context can be either $\{|1\rangle, |D1\rangle, |S1\rangle \}$ or $\{|2\rangle, |D2\rangle, |S2\rangle \}$. The two relations that correspond to Eq.~(\ref{eq:D1D2}) are given by
\begin{eqnarray}
\label{eq:f3vD}
\langle f|3\rangle &=& \langle f|D1\rangle\langle D1|3\rangle,
\nonumber \\
\langle f|3\rangle &=& \langle f|D2\rangle\langle D2|3\rangle.
\end{eqnarray}
An additional relation between $|3\rangle$ and $|f\rangle$ can be obtained by using Eq.~(\ref{eq:f4}) to express the inner product of $|f\rangle$ with itself as
\begin{eqnarray}
\label{eq:D1f}
|\langle f|D1\rangle|^{2}&+&|\langle f|D2\rangle|^{2}-|\langle f|3\rangle|^{2}=1.
\end{eqnarray}
Using Eq.~(\ref{eq:f3vD}) to eliminate the inner products $\langle f|D1\rangle$ and $\langle f|D2\rangle$ from this equation results in an expression for the absolute value of the inner product $\langle f|3\rangle$ that depends only on the inner products $\langle D1|3\rangle$ and $\langle D2|3\rangle$,
\begin{equation}
\label{eq:3f}
|\langle f|3\rangle|^{2}=\frac{|\langle D1|3\rangle|^{2}|\langle D2|3\rangle|^{2}}{1-(1-|\langle D1|3\rangle|^{2})(1-|\langle D2|3\rangle|^{2})}.
\end{equation}
The probabilities of the outcome $|f\rangle$ when the probability of finding $|3\rangle$ is one is therefore completely determined by the probabilities of finding $|D1\rangle$ and $|D2\rangle$ under the same circumstances.

Both the inner products $\langle3|N_{\mathrm f}\rangle$ and $\langle f|3\rangle$ can be determined from the inner products $\langle D1|3\rangle$ and $\langle D2|3\rangle$. According to Eq.~(\ref{eq:Nfnegative}), we can now express the absolute value of the inner product of $|N_{\mathrm f}\rangle$ by taking the product of these two values,
\begin{widetext}
\begin{equation}
\label{eq:main}
|\langle f|N_{\mathrm f}\rangle|^{2}=\frac{|\langle D1|3\rangle|^{2}|\langle D2|3\rangle|^{2}}{1-|\langle D1|3\rangle|^{2}|\langle D2|3\rangle|^{2}}\left(\frac{(1-|\langle D1|3\rangle|^{2})(1-|\langle D2|3\rangle|^{2})}{1-(1-|\langle D1|3\rangle|^{2})(1-|\langle D2|3\rangle|^{2})}\right).
\end{equation}
\end{widetext}
This is the central result of our analysis. It shows that the well-known Hardy-like violation of noncontextuality can be traced back to the fundamental relation between the outcomes $|D1\rangle$ and $|D2\rangle$ that have a probability of zero in the paradox. In the conventional formulation of the paradox \cite{Har92,Har93,Bos97}, the outcome $|N_{\mathrm f}\rangle$ represents the initial state and the outcome $|f\rangle$ represents the outcome that violates the noncontextual expectations. $|\langle f|N_{\mathrm f}\rangle|^{2}$ thus represents the probability of obtaining a paradoxical result of $|f\rangle$ when the probabilities of finding $|D1\rangle$ and $|D2\rangle$ are both zero. Our analysis shows that this probability is completely determined by the probabilities of finding $|D1\rangle$ and $|D2\rangle$ when the outcome $|3\rangle$ has a probability of 1. Quantum contextuality establishes a fundamental relation between these completely different combinations of state preparation and measurement. The relation given in Eq.~(\ref{eq:main}) thus shows that the statistical paradoxes of quantum contextuality can be traced back to more compact elementary relations between only a few contexts, as given by Eq.~(\ref{eq:D1D2}) and Eq.~(\ref{eq:f4}).

The contextuality diagram in Fig.~\ref{fig2} indicates only the orthogonality relations between the measurement outcomes that share a context. Inner products between the Hilbert space vectors representing the outcomes describe a quantitative relation between different contexts, and these quantitative relations define the probabilities of obtaining results that seem to contradict noncontextual logic. Eq.~(\ref{eq:main}) shows that the relations between $|D1\rangle$, $|D2 \rangle$ and $|3\rangle$ are sufficient to determine the probability of observing the seemingly paradoxical outcome $|f\rangle$ for the initial state $| N_{\mathrm f}\rangle$. This relation is by no means obvious. Hardy-like paradoxes are based on the idea that direct contradictions between different measurement contexts are best obtained by finding different outcome probabilities of zero for a given initial state \cite{Har92,Har93,Bos97}. As a result, previous investigations may have overlooked the importance of the relation between the different measurements which do not depend on the choice of an initial state. In the present analysis, the initial state $|N_{\mathrm f}\rangle$ is defined as a measurement outcome that shares a context with both $|D1\rangle$ and $|D2\rangle$. The precise form of $|N_{\mathrm f}\rangle$ is defined by the relation with the central context given by $\langle D1|3\rangle$ and $\langle D2|3\rangle$.

It should be emphasized that the probability $|\langle f|N_{\mathrm f}\rangle|^{2}$ is widely used as a characterization of Hardy-like paradoxes in three dimensional Hilbert spaces. The oldest version of this class of paradoxes is probably the three-box paradox, as explained in~\cite{Lei05}. It is well known that a maximal probability of $|\langle f|N_{\mathrm f}\rangle|^{2}=1/9$ is obtained for an initial state $|N_{\mathrm f}\rangle$ that is an equal superposition of $|1\rangle$, $|2\rangle$, and $|3\rangle$. We can confirm this result by noting that $(a)$ Eq.~(\ref{eq:squareNf}) shows that $|N_{\mathrm f}\rangle$ is an equal superposition of $|1\rangle$, $|2\rangle$, and $|3\rangle$ if and only if $|\langle D1|3\rangle|^2=1/2$ and $|\langle D2|3\rangle|^2=1/2$, and $(b)$ $|\langle f|N_{\mathrm f}\rangle|^{2}=1/9$ is the maximal value obtained from Eq.~(\ref{eq:main}) for values of $|\langle D1|3\rangle|^2=1/2$ and $|\langle D2|3 \rangle|^2=1/2$. The present analysis thus adds the interesting detail that the maximal probability of obtaining a paradoxical measurement outcome in the Hardy-like paradox of three-dimensional Hilbert spaces is observed when the two outcomes that have zero probability are equal superpositions of $|1\rangle$ and $|3\rangle$ for $|D2\rangle$ and $|2\rangle$ and $|3\rangle$ for $|D1\rangle$. In addition to this insight into the most compact Hardy-like paradoxes in three-dimensional Hilbert spaces, the present analysis can also be applied in larger Hilbert spaces, where the additional states represent measurement contexts with all of the measurement outcomes discussed so far. Such additions may be useful when the physics of the system cannot be described completely by a three-dimensional Hilbert space, and this is definitely the case for a product state of two systems. In the following, we therefore discuss the application of our analysis to a system of two separate systems described by the product space of two two-dimensional Hilbert spaces.

\section{Application to quantum nonlocality}
\label{sec:IV}
In the preceding section we have shown that quantum contextuality defines relations between the inner products of Hilbert space vectors representing different measurement contexts and used these relations to show that Hardy-like paradoxes can be traced back to relations between measurement outcomes that are not observed in the original formulation of the paradox. Eq.~(\ref{eq:main}) shows that the probability of observing the paradoxical outcome $|f\rangle$ when the probabilities of observing $|D1\rangle$ and $|D2\rangle$ are zero is given by the seemingly unrelated probabilities of observing $|D1\rangle$ and $| D2 \rangle$ when the probability of finding $|3\rangle$ is one. Since the original paradox of Hardy~\cite{Har92,Har93} was formulated to demonstrate quantum nonlocality, we will now show that only minimal modifications are needed to apply the results of the previous section to entangled quantum systems.

\begin{figure}[ht]
\begin{picture}(500,200)
\put(50,0){\makebox(130,200){
\scalebox{0.55}[0.55]{
\includegraphics{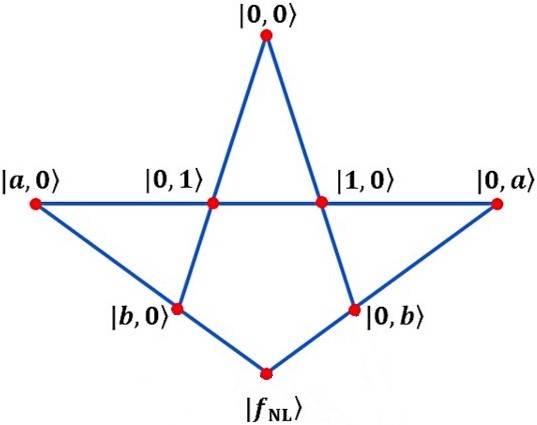}}}}
\end{picture}
\caption{\label{fig3}
Definition of measurement outcomes in the product space of two two-level systems. Seven of the eight measurement outcomes can be replaced with product states. Only the outcome $|f_{\mathrm {NL}}\rangle$ is represented by an entangled state, indicating a nonlocal measurement of a collective property of both systems.
}
\end{figure}

In general, the measurement outcomes in Fig.~\ref{fig2} could refer to any physical system. It is therefore possible to apply them to a product space describing a pair of two-level systems, where each measurement outcome can be identified with a product state of the two systems. For simplicity, we assume that the two systems have identical physical properties. Since the outcome $|3\rangle$ should not distinguish between the two systems, we identify it with the product state $|0,0\rangle$. The outcome $|1\rangle$ is identified with $|0,1\rangle$, and the outcome $|2\rangle$ is identified with $|1,0\rangle$. We now need to find a product state outcome that is a superposition of $|0,0\rangle$ and $|1,0\rangle$ ($|0,1\rangle$) to replace $|D1\rangle$ ($|D2\rangle$). This is also a product state, and we define the local superposition of $|0\rangle$ and $|1\rangle$ that defines it as local outcome $|a\rangle$. The outcome that takes the place of $|D1\rangle$ ($|D2\rangle$) is then given as $|a,0\rangle$ ($|0,a\rangle$). Likewise, $|S1\rangle$ ($|S2\rangle$) is replaced with the local outcomes $|b,0\rangle$ ($|0,b \rangle$). Up to now, all of the outcomes are local. However, the remaining outcome $|f\rangle$ is determined by the condition that it must be a superposition of the three states $|0,0\rangle$, $|0,1\rangle$ and $|1,0\rangle$. In general, $|f\rangle$ will be an entangled state describing a nonlocal measurement. We therefore replace it with $|f_{\mathrm {NL}}\rangle$. Fig.~\ref{fig3} shows the seven local outcomes and the one nonlocal outcome in the same contextuality diagram as the one shown in Fig.~\ref{fig2}, highlighting the fact that the Hilbert space relations between the outcomes are exactly identical to the ones derived above.

Since the only difference between Fig.~\ref{fig3} and Fig.~\ref{fig2} is the change of the labels used to identify the quantum states, no new analysis of the problem is needed and all of the previous results can be applied directly to the product space of two two-level systems. The product state representation allows us to express all inner products using the local relation between $|a\rangle$ and $|0\rangle$. Note that the definition of the outcome $|N_{\mathrm f}\rangle$ is completely unchanged. We can therefore keep the terminology and define it as the outcome that shares a context with both $|a,0\rangle$ and $|0,a\rangle$, and with any addition to the context $\{|0,0\rangle,|0,1\rangle,|1,0\rangle\}$. In the four dimensional Hilbert space of the two two-level systems, this additional state is $|1,1\rangle$. According to Eq.~(\ref{eq:main}), the probability of finding $|f_{\mathrm {NL}}\rangle$ when the probabilities of finding $|a,0\rangle$, $|0,a\rangle$ or $|1,1\rangle$ are all zero can be expressed completely in terms of the probability $|\langle a|0\rangle|^2$ of finding $|a\rangle$ in a single two-level system when the probability of finding $|0\rangle$ is one. The result is given by
\begin{equation}
\label{eq:NL1}
|\langle f_{\mathrm {NL}}|N_{\mathrm f}\rangle|^{2}=\frac{|\langle a|0\rangle|^{4}}{1+|\langle a|0\rangle|^{2}}\left(\frac{1-|\langle a|0\rangle|^{2}}{1-(1-|\langle a|0\rangle|^{2})^{2}}\right).
\end{equation}
This is a nonlocal form of contextuality, since it refers to a measurement outcome $|f_{\mathrm {NL}}\rangle$ that cannot be defined by a product state. Interestingly, local measurement outcomes can be used to define both $|N_{\mathrm f}\rangle$ and $|f_{\mathrm {NL}}\rangle$. However, the relation in Eq.~(\ref{eq:NL1}) is a genuine nonlocal contextuality, different from the original Hardy paradox. To recover the relation between local contextuality and quantum nonlocality discussed in~\cite{Cab10,Liu16}, it is necessary to find a product state to replace the measurement outcome $|f_{\mathrm {NL}}\rangle$, so that the origin of the nonlocality is exclusively associated with the outcome $|N_{\mathrm f}\rangle$ that plays the role of the initial state in the conventional formulation of demonstrations of quantum nonlocality.

\begin{figure}[ht]
\begin{picture}(500,200)
\put(50,0){\makebox(130,200){
\scalebox{0.55}[0.55]{
\includegraphics{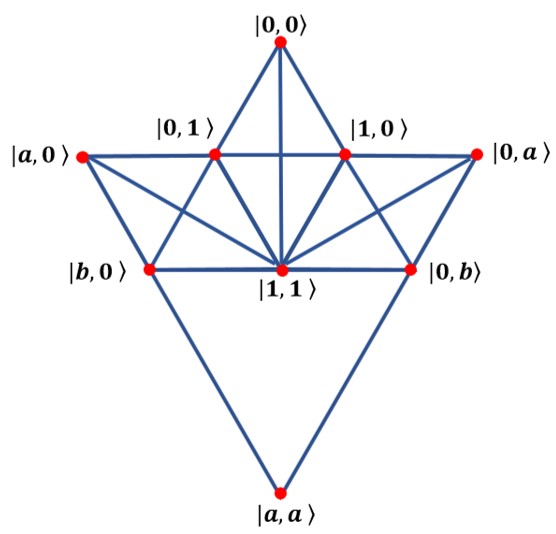}}}}
\end{picture}
\caption{\label{fig4}
Diagram including the outcome $|1,1\rangle$ and the modified final outcome $|a,a\rangle$. Based on the initial definitions of the outcomes, $|1,1\rangle$ is orthogonal to seven of the eight other states in the diagram.
}
\end{figure}

The reason why $|f_{\mathrm {NL}} \rangle$ is nonlocal is the constraint that it must share a context with $|1,1\rangle$. If this constraint is lifted, it is possible to replace $|f_{\mathrm {NL}} \rangle$ with a product state that shares a context with $|b,0\rangle$ and $|0,b\rangle$. The obvious solution is the state $|a,a\rangle$. Fig.~\ref{fig4} shows the contextuality diagram including both $|1,1\rangle$ and $|a,a\rangle$. Since the state $|a,a\rangle$ is orthogonal to both $|b,0\rangle$ and $|0,b\rangle$, just like the state $|f_{\mathrm {NL}}\rangle$, its projection into the three dimensional subspace of $\{|0,0\rangle, |0,1\rangle, |1,0\rangle\}$ must be given by $|f_{\mathrm {NL}}\rangle$. It is therefore possible to express the state $|a,a\rangle$ as a superposition of $|f_{\mathrm NL}\rangle$ and $|1,1\rangle$,
\begin{equation}
\label{eq:aa}
|a,a\rangle=|f_{\mathrm {NL}}\rangle\langle f_{\mathrm {NL}}|a,a\rangle+|1,1\rangle\langle1,1|a,a\rangle.
\end{equation}
Note that it is not at all trivial that a product state can be written as a superposition of a specific entangled state and an additional product state. The orthogonality relations given by the diagrams in Figs.~\ref{fig3} and \ref{fig4} are needed to establish the necessary relation between the entangled state $| f_{\mathrm {NL}}\rangle$ and the product state $|a,a\rangle$ given by Eq.~(\ref{eq:aa}).

Eq.~(\ref{eq:aa}) shows that the probability of finding the measurement outcome $|f_{\mathrm {NL}}\rangle$ when the probability of $|a,a\rangle$ is one can be given by
\begin{eqnarray}
\label{eq:faa}
|\langle f_{\mathrm {NL}}|a,a\rangle|^{2}&=&1-|\langle 1,1|a,a\rangle|^{2}
\nonumber \\
&=& 1-(1-|\langle0|a\rangle|^{2})^{2}.
\end{eqnarray}
Since the state $|N_{\mathrm f}\rangle$ is orthogonal to $|1,1\rangle$, only the component $|f_{\mathrm {NL}}\rangle$ in the superposition shown in Eq.~(\ref{eq:aa}) contributes to the inner product of $|a,a\rangle$ and $|N_{\mathrm f}\rangle$,
\begin{equation}
\label{eq:NNLaa}
\langle a,a|N_{\mathrm {f}}\rangle=\langle a,a|f_{\mathrm {NL}}\rangle\langle f_{\mathrm {NL}}|N_{\mathrm {f}}\rangle.
\end{equation}
The probability of finding $|a,a\rangle$ in $|N_{\mathrm f}\rangle$ is therefore given by the product of the probabilities shown in Eqs.~(\ref{eq:NL1}) and (\ref{eq:faa}). The result determines the probability of finding the outcome $|a,a\rangle$ when the probabilities of $|a,0\rangle$, $|0,a\rangle$ and $|1,1 \rangle$ are all zero as a function of the local relation between contexts given by the inner product $\langle a|0\rangle$,
\begin{equation}
\label{eq:NNLaa2}
|\langle a,a|N_{\mathrm {f}}\rangle|^{2}=|\langle a|0\rangle|^{4}\left(\frac{1-|\langle a|0\rangle|^{2}}{1+|\langle a|0\rangle|^{2}}\right).
\end{equation}
This result not only confirms that the nonlocality of Hardy's paradox originates from the local contextuality relation, it also provides a precise quantitative relation between the local inner product $\langle a|0\rangle$ and the probability of observing the outcome associated with the quantum nonlocality. Specifically, the probability of observing $|a,a\rangle$ for an entangled state defined by zero probability for $|a,0\rangle$, $|0,a\rangle$ and $|1,1\rangle$ is uniquely defined by the probability of finding $|a\rangle$ in a single system when the probability of finding $|0\rangle$ in that system is one. It might be worth noting that the specific case of $ |\langle a|0\rangle|^{2}=1/2$ corresponds to the scenario discussed by Frauchiger and Renner, where the probability of finding the outcome $|a,a\rangle$ is $|\langle a,a|N_{\mathrm {f}}\rangle|^{2}=1/12$ \cite{Ji23,Fra18}.

As mentioned above, the relation between quantum nonlocality and contextuality has been widely discussed in the literature~\cite{Cab10,Liu16,Kup23}, so it is not surprising that a quantitative analysis of contextuality has implications for quantum nonlocality. However, it seems to be remarkable that the connection between the basic contextuality relations involving only three Hilbert space dimensions is so easily extended to the minimal four dimensional Hilbert space needed for two spatially separated systems. It should be noted that the most important step was the identification of the measurement outcomes with product states of the separate systems, resulting in a relaxation of the requirement that constrained the measurement outcome $|f\rangle$ to the three dimensional subspace of Hilbert space defined by the initial context. It is also worth noting that the general formula given in Eq.~(\ref{eq:main}) described the relation between $|D1\rangle$ and $|D2\rangle$ in terms of two independent inner products with the intermediate state $|3\rangle$, whereas the symmetry between the two spatially separated systems in the analysis of the present section reduces the input to only a single inner product relating the local outcomes $|a\rangle$ and $|0\rangle$ to each other. Somewhat surprisingly, nonlocality appears to be an especially simple example of quantum contextuality, since all of the constraints added to introduce the mathematical structure of product spaces actually help to simplify the equations. It remains to be seen whether similar simplifications can be obtained when the method developed here is applied to larger networks and multi-partite entanglement.

\section{Conclusion}
\label{sec:V}
We have investigated the manner in which the Hilbert space formalism expresses quantum contextuality by starting from an elementary relation between two measurement outcomes $|D1\rangle$ and $|D2\rangle$ connected by a central context. The paradoxes associated with quantum contextuality can then be explained by introducing additional relations between $|D1\rangle$ and $|D2\rangle$. Our central result shows that the most basic Hardy-like paradox defined in a three-dimensional subspace of Hilbert space can be traced back to the relation between $|D1\rangle$ and $|D2\rangle$ with the outcome $|3\rangle$ in the central context. Significantly, we can show that the probability of obtaining a paradoxical outcome in the Hardy-like paradox is completely determined by measurement probabilities observed in seemingly unrelated measurement scenarios. We conclude that the Hilbert space formalism defines a much richer structure of quantum contextuality than the paradoxes themselves suggest. If measurements are related to specific quantum state preparation procedures that allow the prediction of the associated measurement outcome with certainty, the mathematical description of the relation between different measurement contexts obtains an experimentally testable meaning, and the relation between different measurement contexts can be discussed as part of the logical consistency of quantum statistics. In the present paper, we have developed the method of relating different measurement contexts to each other and applied it to the most compact systems that exhibit quantum contextuality. We emphasize that these results are not limited to three-dimensional Hilbert spaces, as shown by the extension of the initial result to the product space of two two-level systems. Hardy's proof of quantum nonlocality can be obtained by a small modification of the original contextual network, illustrating the close relation between nonlocality and contextuality and identifying local measurement probabilities as the origin of nonlocality in the global statistics. The new results presented in this paper show that contextuality is a natural feature of a formalism that relates different measurement contexts by superpositions and inner products of Hilbert space vectors, resulting in a wide range of relations between measurement probabilities observed in completely different measurement scenarios. The method introduced here makes these fundamental relations between different measurement contexts more accessible and may result in new fundamental insights into the nature of quantum phenomena. In particular, it might be possible to systematically derive the essential nonclassical relations between the statistics of incompatible measurements by quantitative relations between the inner products of the Hilbert space vectors derived from elementary relations between different measurement contexts.

\section*{Acknowledgment}
We would like to thank Dr. J. R. Hance for helpful comments. This work was supported by JST SPRING, Grant Number JPMJSP2132.

\bibliographystyle{unsrturl}
\bibliography{ref.bib}

\end{document}